**Detecting the Major Charge-Carrier Scattering Mechanism in Graphene Antidot Lattices**

Dongchao Xu[1], Shuang Tang[2]**, Xu Du[3], and Qing Hao[1]*

[1] *Department of Aerospace and Mechanical Engineering, University of Arizona, Tucson, AZ 85721, USA*

[2] *College of Engineering, State University of New York, Polytechnic Institute, Albany, NY, 12203, USA*

[3] *Department of Physics, Stony Brook University, Stony Brook, NY 11790, USA*

Corresponding authors: *qinghao@email.arizona.edu, ** tangs1@sunyit.edu

**ABSTRACT:** Charge carrier scattering is critical to the electrical properties of two-dimensional materials such as graphene, transition metal dichalcogenide monolayers, black phosphorene, and tellurene. Beyond pristine two-dimensional materials, further tailored properties can be achieved by nanoporous patterns such as nano- or atomic-scale pores (antidots) across the material. As one example, structure-dependent electrical/optical properties for graphene antidot lattices (GALs) have been studied in recent years. However, detailed charge carrier scattering mechanism is still not fully understood, which hinders the future improvement and potential applications of such metamaterials. In this paper, the energy sensitivity of charge-carrier scattering and thus the dominant scattering mechanisms are revealed for GALs by analyzing the maximum Seebeck coefficient with a tuned gate voltage and thus shifted Fermi levels. It shows that the scattering from pore-edge-trapped charges is dominant, especially at elevated temperatures. For thermoelectric interests, the gate-voltage-dependent power factor of different GAL samples are measured as high as 554 μW/cm·K$^2$ at 400 K for a GAL with the square pattern. Such a high power factor is improved by more than one order of magnitude from the values for the state-of-the-art bulk thermoelectric materials. With their high thermal conductivities and power factors, these GALs can be well suitable for "active coolers" within electronic devices, where heat generated at the hot spot can be removed with both passive heat conduction and active Peltier cooling.

Understanding the scattering mechanisms of charge carriers is critical to research fields such as electronics, thermoelectrics, and mechatronics. In general, the charge carrier transport is governed by the Boltzmann transport equation[1,2]. In such modeling, the scattering rate of charge carriers is one major input parameter, where charge carriers are affected by polar/nonpolar acoustic and optical phonons[3], ionized impurities[4], alloy disorder[5] and other defects[1]. For two-dimensional (2D) materials, the scattering processes are often divided into short- and long-range scattering processes. The short-range scattering processes normally include point-contact scattering[6], vacancy scattering[7], while long-range scattering processes include Coulomb scattering[8], thermal ripple scattering[9] and ionized center scattering[10,11]. For patterned 2D materials such as graphene nanoribbons, the edge roughness scattering[12] should be further considered as short-range scattering. Generally, each scattering mechanism can be characterized by a scattering rate $\tau^{-1}$. The carriers' energy sensitivity ($j$) to scattering (CEStS) can be described as, $j = -\left[d(\tau^{-1})/(\tau^{-1})\right]/[d\varepsilon/\varepsilon] = d\left[\ln(\tau)\right]/d\left[\ln\varepsilon\right]$, where $\varepsilon$ is the carrier's energy referring to the edge of the corresponding valley. Different scattering mechanisms have their own CEStS values[13]. In the theoretical models for individual scattering mechanisms, involved parameters can be fitted with the measured electrical conductivity ($\sigma$)[14]. Experimentally, an averaged scattering rate can be estimated by observing the relaxation time of photon-electron interaction using an ultra-fast laser[15, 16]. However, the co-existence of many scattering mechanisms often adds large uncertainties to these studies. As an alternative approach, Tang recently proposed to find the CEStS by measuring the maximum Seebeck coefficient (*S*) under a shifted Fermi level[13]. By comparing the extracted effective CEStS with known CEStS values for existing scattering mechanisms, the major scattering mechanism can be inferred. For 2D materials, such Seebeck coefficient measurements can be easily carried out by tuning the Fermi level with a gate voltage[17, 18]. The strong gate effect of 2D materials also ensure the observation of maximum Seebeck values within the typical voltage range of a power supply[19].

In this work, Tang's model[13] is employed to better understand the electron scattering within monolayer graphene patterned with periodic nanoscale pores (antidots), known as graphene antidot lattices (GALs)[20-22] and sometimes graphene nanomesh[23]. Viewed as networked nanoribbons, GALs employ the narrow neck width (~10 nm or less) between antidots to confine charge carriers and thus open a geometry-dependent band gap within otherwise gapless graphene. Compared with nanoribbons[12, 24, 25], GALs can carry a much higher electrical current for applications in electronic devices. Beyond field effect transistors, GALs have been suggested for their wide applications in gas sensing[26], thermoelectric power generation and cooling[27, 28], optoelectronic devices[29], magnetics[30, 31], spintronics[32, 33] and waveguides[34]. Along this line, numerous simulations and measurements have been carried out to understand the structure-dependent electrical properties of GALs. Across the narrow neck width between antidots, ballistic electron transport is suggested in some experimental studies[21, 22]. However, the dominant scattering mechanism is still not fully understood. Although charge carriers are likely to be frequently scattered by pore edges, the exact pore-edge atomic configuration (e.g., armchair or zigzag edges) and charges trapped on the pore edges may both affect the energy-dependent charge carrier scattering. For comparable graphene nanoribbons, analytical models have been particularly developed for electron scattering within armchair graphene nanoribbons[35]. With such complexity, simply diffusive and elastic charge carrier scattering cannot be assumed at pore edges.

In this work, the dominant electron scattering mechanism within representative GALs on a $SiO_2$/Si substrate is revealed with gate-tuned Seebeck coefficient measurements from 82 to 400 K. Both square and hexagonal arrays of nanopores are fabricated. By analyzing the maximum Seebeck coefficient values under a tuned gate voltage, it is found that the electrons are mainly scattered by the pore-edge-trapped charges, particularly at elevated temperatures. For thermoelectric interests, the best power factor $S^2\sigma$ under an applied gate voltage is found in a n-type GAL with the square pattern. The maximum $S^2\sigma$ value can be as high as 292 and 554 μW/cm·$K^2$ at 300 and 400 K, respectively. For on-chip cooling applications, such values are already much higher than the room-temperature power factor of ~45 μW/cm·$K^2$ for the state-of-the-arts bulk BiSbTe alloys[36]. The demonstrated approach can be applied to general 2D materials and their antidot lattices (e.g., black phosphorene[37], silicene[38]) in future studies.

Monolayer GALs with hexagonal or square arrays of nanopores (antidots) were fabricated for the proposed study. Nanofabricated thermal and/or electrical probes were deposited onto each GAL to ensure intimate thermal/electrical contacts between these probes and the sample. The electrical conductivity and the Seebeck coefficient were measured under different gate voltages applied from the degenerately doped Si substrate.

Figure 1 displays the scanning electron microscope (SEM) images of two representative samples. The center-to-center distance between adjacent pores, also called pitch, are fixed as 30 nm. The neck width is 14.2±1 nm for both patterns. All nanopores are accurately defined by electron beam lithography (EBL). The original monolayer graphene is purchased from Graphenea Inc. These graphene samples are grown on a Cu foil through chemical vapor deposition (CVD) and then transferred onto a SiO$_2$/Si substrate. In a recent work also using CVD-grown graphene, GALs are fabricated with block copolymer (BCP) self-assembly as the mask for reactive ion etching[27]. Only roughly hexagonal patterns can be defined by the BCP film. In contrast, the use of EBL in our study allows better structure controls for fundamental studies.

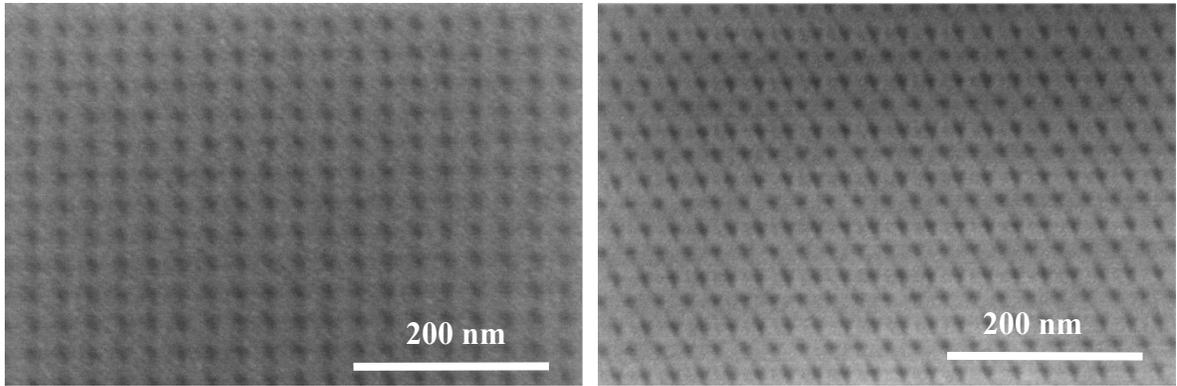

**FIGURE 1**. SEM images of GALs with a (a) square and (b) hexagonal array of nanopores. Scale bars are 200 nm for both images.

Figure 2 (a) presents the measurement scheme and Figure 2 (b) further shows the SEM image of a real setup. More details are shown in the inset of Figure 2 (b). In addition, a gate voltage can also be applied from the back of the degenerately doped Si substrate to tune the Fermi level and thus the electrical properties of the GALs. The electrical conductivity $\sigma$ is measured with standard four-probe technique. Two outer electrodes (source and drain) are used for 1-kHz AC current injection, while the inner two electrodes are used to measure the voltage drop. For the convenience of $\sigma$ measurements, the CVD graphene is trimmed as a strip with patterned periodic nanopores.

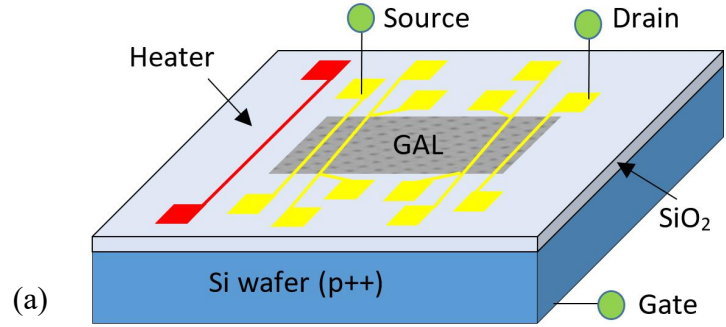

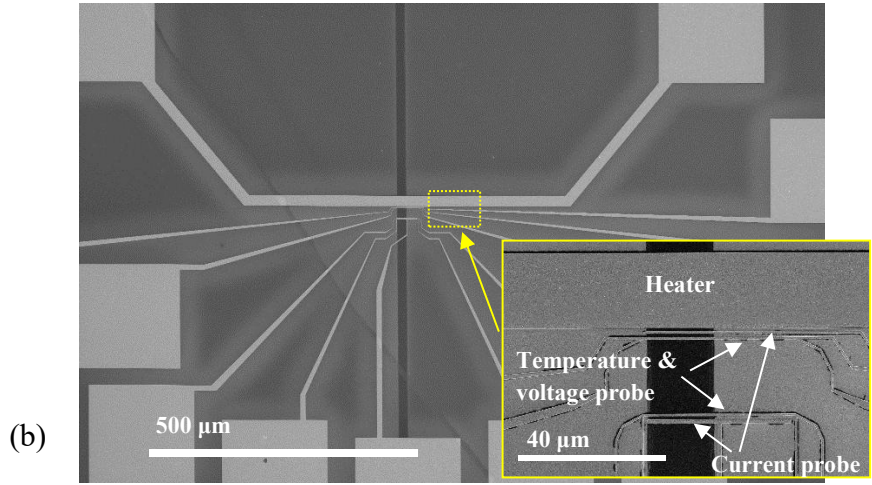

**FIGURE 2**. (a) Measurement scheme. (b) SEM images of the measurement setup, with the inset as the GAL region to be measured.

The same setup is also used to measure the Seebeck coefficient ($S$) simultaneously. Using a micro-fabricated heater near the left side of the sample, a temperature gradient can be created. To avoid current leakage from the heater to the sample, the graphene layer between the heater and its nearest electrode for the current injection is cut with a focused ion beam (FIB). Two inner electrodes are now used as thermometers, with the local temperature $T$ obtained from their temperature-dependent electrical resistance. The metal-line heater is much wider than the measured GAL strip so that the temperature variation along the two metal-line thermometers can be neglected. A similar setup has been used in the Seebeck measurements of nanoporous Si films[39]. In addition to the temperature gradient measurements, the same two inner probes are also used to measure the voltage drop $\Delta V$. With the temperature difference $\Delta T$ between two metal-line thermometers, the slope of the $\Delta T$-$\Delta V$ curve is extracted to compute $S=-d\Delta V/d\Delta T+S_{Au}$, in which $S_{Au}$ is the compensated Seebeck coefficient of the Au voltage probes[36].

At different temperatures, the CEStS (*j*) of each sample are extracted using Tang's method through the maximum $|S|$ values of both p and n types, which is measured by shifting the Fermi level with an applied gate voltage. The CEStS values can then be used as the indicator of the scattering mechanism(s).

In order to obtain the CEStS, the carrier's energy sensitivity to transport (CEStT) should be computed first. Here the transport function is given as $\Xi = \tau D(E)\langle v^2 \rangle_E$, where $D(E)$ is the density of states at E, *v* is the group velocity and $\langle . \rangle_E$ stands for the mean value at the constant energy surface. The CEStT can be defined as $s=d\ln(\Xi)/d\ln(E)$, which means the ratio between percent change in transport and percent change in carrier energy. The relation between *s* and *j*, is then $j = s - d\ln(D(E)\langle v^2 \rangle_E)/d\ln(E)$. Hence, j can be calculated once the *s* value and the electronic band structure are known.

Tang shows that each conduction (valence) band valley that contribute to the transport will result in a negative (positive) peak/kink value of the Seebeck coefficient, while the Fermi level is tuned through a gate voltage[13]. The *i*th peak/kink value $S_m^i$, form a near-linear relation with the energy sensitivity corresponding to the *i*th band valley, i.e.,

$$S_m^{[i]} \approx s^{[i]} \cdot \frac{k_B}{q} \cdot a^{[i]} + S_0^{[i]}, \tag{1}$$

where $a^{[i]}$ and $S_0^{[i]}$ are quantities specified by the band gap and temperature. Therefore, the energy sensitivity of transport $s$ and thus $j$ at a certain temperature can be calculated by measuring the $S_m^i$ values of a system. The value of $j$ stands for a mechanism-specific energy sensitivity for a single scattering mechanism associated with a specific band valley, and for a weighted average of all existing scattering mechanisms for cases with coexistence of multiple scattering mechanisms.

For the carriers near the band edges in semiconducting GALs, $j=0$ is anticipated for acoustic phonon scattering and inelastic scattering. Short range disorder scattering typically has $j<0$, such as $j$ for point contact, point defects and vacancies scattering. In addition, typical elastic boundary scattering at pore edges also has negative $j$. For this scattering, the corresponding relaxation time is determined by the characteristic length of the nanoporous structure[40] divided by the group velocity $v_g$ of charge carriers. Because $v_g$ increases with the energy $\varepsilon$, the relaxation time $\tau$ decreases with $\varepsilon$ and leads to a negative $j = d[\ln(\tau)]/d[\ln \varepsilon]$. The rest scattering mechanisms are long range and have $j>0$, including Coulomb interaction scattering and thermal ripples scattering[13].

As the Arrhenius plot, Figure 3 shows the temperature-dependent electrical conductivity $\sigma$ corresponding to the OFF conductance. With these temperature-dependent $\sigma$ values, the bandgap $E_g$ can be extracted by fitting[27, 41]

$$\sigma = \sigma_0 \exp[-E_g/(2k_B T)], \tag{2}$$

where $\sigma_0$, $k_B$ and $T$ are a fitted constant, the Boltzmann constant and absolute temperature, respectively. Equation (2) attributes the temperature dependence of $\sigma$ to thermally activated charge carriers. A band gap $E_g \approx 47.4$ meV is extracted for the square pattern, in comparison to $E_g \approx 53$ meV for the hexagonal pattern. With a given pitch $P$=30 nm and averaged pore diameter $d \approx 15.8$ nm, the hexagonal pattern has stronger quantum confinement for charge carriers and thus a larger $E_g$ with a consistent neck width $P-d$ between adjacent nanopores, whereas the square pattern has an expanded neck width $\sqrt{2}P-d$ between the second-nearest-neighbor nanopores. Other than the neck width, the hexagonal pattern also has a 17% smaller characteristic length. Based on the mean beam length used for radiation, the characteristic length for a periodic 2D nanoporous structure is proportional to the solid-region area of one period divided by the pore perimeter[40, 42]. A smaller characteristic length indicates more influence from nanopores. These bandgap values are comparable to reported $E_g \approx 60$ meV for a single-layer hexagonal GAL with a neck width $n$=12 nm [27]. This reported band gap is estimated with the same technique.

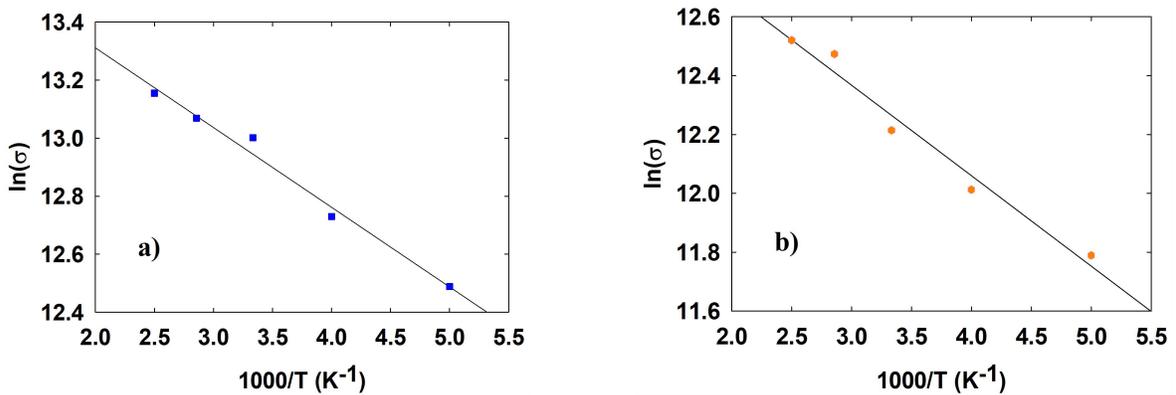

**FIGURE 3**. Temperature-dependent electrical conductivity under zero gate voltage for (a) square and (b) hexagonal patterns.

Figures 4 (a) and (b) present the gate-voltage-tuned electrical conductivity $\sigma$ and Seebeck coefficient $S$ at 300 K. Away from the charge-neutrality point, a negative gate voltage can induce holes within the material as the "hole side." On the right side of both figures, a positive gate voltage can lead the sample into the "electron side." In the vicinity of the charge-neutrality point, the Seebeck coefficients of co-existing electrons and holes can largely cancel out. For GALs, n-type doping is observed here and is consistent with one previous study[27]. Again the hexagonal pattern leads to a smaller $\sigma$ because of its smaller characteristic length and thus stronger charge carrier scattering due to nanopore boundaries and pore-edge-trapped charges. All measurement results are further compared to those reported for uncut pristine single-crystal graphene on a $SiO_2$ or hexagonal boron nitride (h-BN) substrate, measured with a similar setup[43]. For these undoped single-crystal samples, the charge neutrality point lies approximately at $V_g$=0 V. In this case, $S$ and $S^2\sigma$ are both zero without an applied gate voltage. With antidots, reduced electrical conductivities are observed in GALs, whereas Seebeck coefficients are generally enhanced with band gaps opened in GALs. Compared to the existing electrical measurements of hexagonal single- and bi-layer GALs using a commercial setup with pressure contacts (TEP-600, Seepel Instrument, Korea)[27], the electrical properties measured in this work are more accurate due to the much better thermal contacts between the sample and deposited metallic probes. Although electrical contacts may not be a concern for $\sigma$ measurements, the measurements of temperature difference $\Delta T$ across the GAL can be largely affected by the thermal contact between the sample and the thermal probes. When the thermal contacts were not good, $S = -\Delta V/\Delta T$ can be underestimated due to overestimated $\Delta T$. As well acknowledged in the literature, , such $S$ underestimation[44] may not be consistent due to possibly improved thermal contacts at elevated temperatures. In this aspect, metal deposition can ensure good thermal contact and is also used for thermal measurements of graphene using a microdevice[45]. In the reported data for single-layer hexagonal GALs on a SiO$_2$/Si substrate, the room-temperature $S$ changes from 5±2 µV/K for pristine CVD graphene to -12±5 µV/K for a hexagonal GAL with neck width $n$=12 nm and an estimated band gap $E_g \approx 60$ meV [27]. On the contrary, a dramatically improved $S = -190 \pm 80$ µV/K is found for a bilayer GAL with neck width $n$=8 nm and a much smaller $E_g \approx 25$ meV. This contradiction may be attributed to the large uncertainties due to poor thermal contacts in the Seebeck coefficient measurements. In this work, more accurate measurements suggest significant Seebeck enhancement and the gate-voltage-dependent $S$ is further measured to better understand the electron transport.

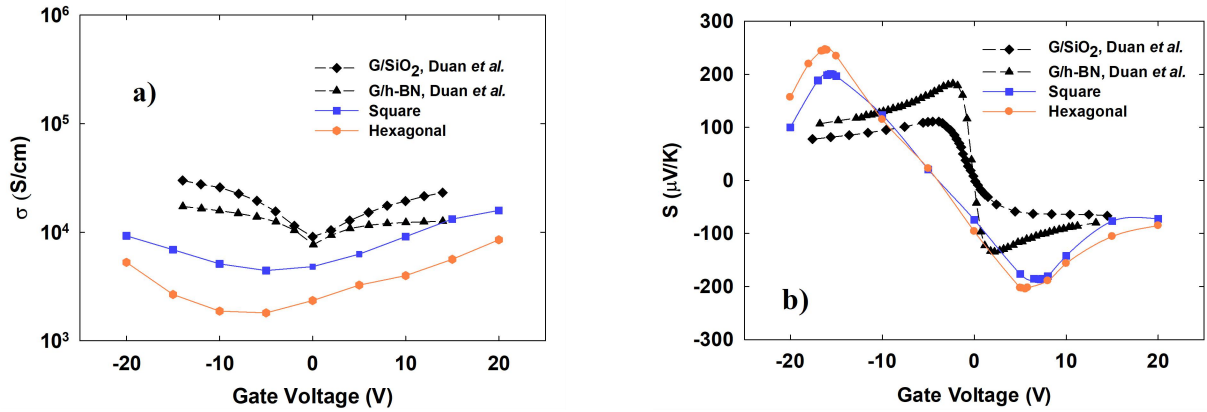

**FIGURE 4.** Gate-voltage-tuned (a) electrical conductivities and (b) Seebeck coefficients at room temperature. All data are further compared with pristine single-crystal graphene on h-BN and SiO$_2$ substrates[43].

Using the maximum $|S|$ for both the p and n types, the CEStS of GAL samples with square and hexagonal patterns are further calculated using an open-access code developed by Tang[13] (Figure 5). The extracted $j$ in Figure 5 reflects a statistical measure of the average values of $j$ for all existing scattering sources. For electrons and holes, their $j$ values diverge from the averaged value within 10% and the averaged $j$ value is plotted here. Apparently, the long-range scattering sources ($j>0$) are important at low temperatures and become dominant at high temperatures. The scattering mechanisms include scattering by ionized center, thermal ripples and/or trapped charges at pore edges. The scattering by ionized centers[10, 11] and thermal ripples[9] has been well studied for graphene, giving relatively long mean free paths (MFPs) for charge carriers. Besides these, analytical modeling of trapped charge scattering at the pore edges of general porous 2D materials and thin films has also been developed[40]. The trapped charges lead to a cylindrical electric field around each pore to scatter nearby charge carriers. When the spacing between adjacent nanopores is decreased below the MFPs of charge carriers in bulk graphene, the scattering by ionized centers and thermal ripples is negligible[21, 22] so that scattering by pore-edge-trapped charges becomes dominant. With a smaller characteristic length and stronger influence from pore-edge-trapped charges, the hexagonal pattern yields a larger $j$ value for a GAL.

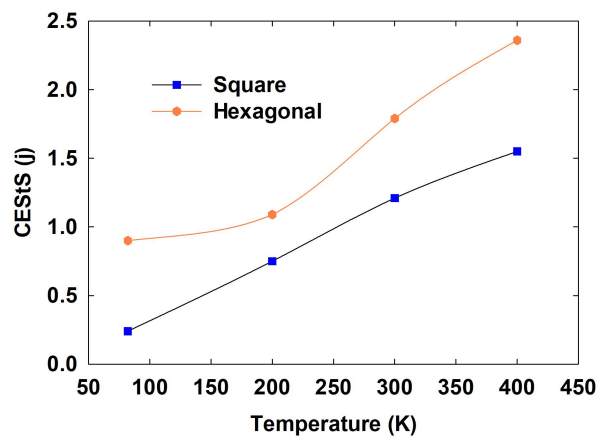

**FIGURE 5.** The CEStS ($j$) of GALs with (a) a square array and (b) a hexagonal array of nanopores.

Without the applied gate voltage, temperature-dependent electrical conductivity $\sigma$, Seebeck coefficient $S$, and the corresponding power factor $S^2\sigma$ are displayed in Figures 6 (a)-(c), respectively. The measurement data for the pristine CVD graphene, cut as a 17-µm-wide microribbon, is also plotted for comparison. Slight p-type doping is found and can be attributed to oxygen dangling bonds introduced by oxygen plasma etching[46]. Despite some p-type doping for this microribbon, its $S^2\sigma$ is much smaller compared with GALs. In Figure 6(c), all power factors are also compared with the state-of-the-art nanostructured bulk BiSbTe (nano BiSbTe) alloys, as the best room-temperature thermoelectric material[47]. Although pristine graphene has zero band gap and a poor $S$, GALs have significantly improved $S$ and thus $S^2\sigma$. The power factor of GALs monotonously increases at elevated temperatures and starts to exceed that for nano BiSbTe alloys above 347 K. The overall trend for temperature-dependent power factor is consistent with the measurements on hexagonal GALs[27]. One major difference between GALs and nano BiSbTe alloys (bulk $E_g \approx 0.3$ eV) lies in that the detrimental bipolar conduction is not observed in GALs though GALs have a very small $E_g$ here. The bipolar conduction origins from thermally activated minority carriers at elevated temperatures, which will cancel out the Seebeck coefficient of majority carriers[36]. When bipolar conduction occurs, $S$ and $S^2\sigma$ usually start to decay but this trend is not found for GALs. The remarkable suppression of minority carrier transport can be attributed to the strong scattering of minority charge carriers by the pore-edge electric field[40]. Similar mechanisms have been proposed for polycrystalline bulk materials with an interfacial energy barrier to filter out more minority carriers than majority carriers[47, 48].

The thermal conductivity $k$ is not measured here to evaluate the thermoelectric figure of merit (ZT), where ZT is defined as $ZT = S^2\sigma T/k$, with $T$ as the absolute temperature[36]. However, $k$ is anticipated to be high according to the existing two-laser Raman thermometry measurements on suspended hexagonal monolayer GALs, which gives $k \approx 337\pm26$ W/m·K for neck width $n$=12 nm, and $k \approx 579\pm42$ W/m·K for $n$=16 nm close to the room temperature[27]. Such a high thermal conductivity can be suppressed by the presence of a substrate[49] but the reduction is anticipated to be limited. The resulting low ZT is not desirable for refrigeration applications. Typical thermoelectric refrigerators usually cool down an object to a temperature lower than that for the heat-rejection junction. In this case, a low $k$ of the thermoelectric material is preferred to block the backward heat conduction and ZT is used to evaluate the material effectiveness. However, this is not the case for electronic cooling in which the hot spot to be cooled always has the highest temperature within the device[50]. In this case, a high $k$ (for passive cooling via heat conduction) combined with a high $S^2\sigma$ (active cooling) are required to better dissipate the heat from the hot spot. With their extremely high power factors and thermal conductivities, GALs can be ideal for such "active coolers" in electronic devices.

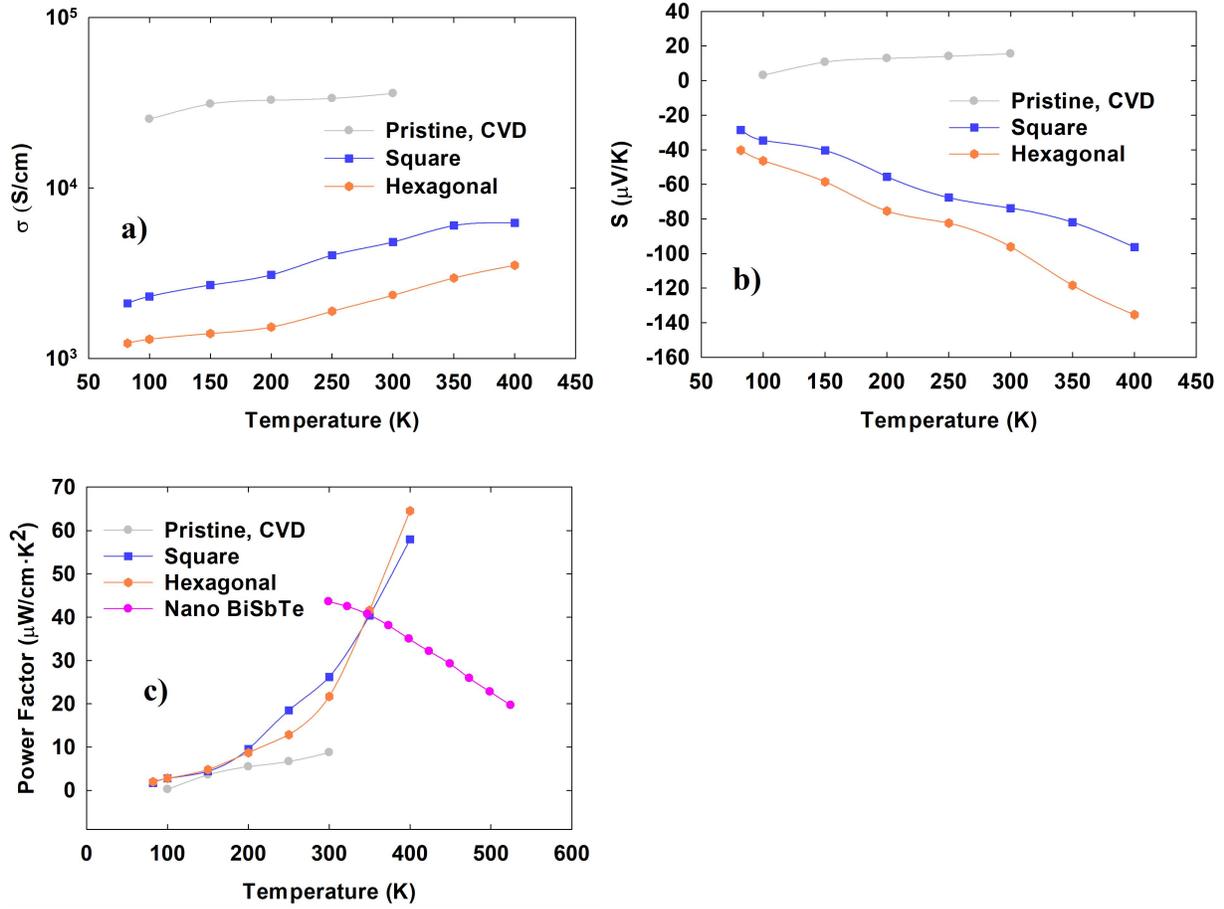

**FIGURE 6**. Temperature-dependent (a) electrical conductivity, (b) Seebeck coefficient, and (c) power factor of measured GALs and pristine CVD graphene cut as a microribbon. The state-of-the-art nanostructured bulk BiSbTe alloys[47] are further compared in (c). No gate voltage is applied to GALs here.

By tuning the gate voltage at each temperature, the power factor can be largely enhanced. Figures 7 (a)-(d) shows the optimized gate voltage $V_g$ to maximize $|S|$, the corresponding $\sigma$, $S$, and $S^2\sigma$ at different temperatures, respectively. At $V_g$ around 5 V, the GALs show n-type properties (i.e., electron side), while p-type properties are shown at $V_g$ around -15 V (i.e., hole side). For the same GAL, the power factors of both types are very close. Between the two GALs, the square pattern exceeds the hexagonal pattern for the power factor enhanced by $V_g$. The results are compared with that measured for pristine single-crystal graphene on h-BN and SiO$_2$ substrates, with a fixed gate voltage and thus constant carrier concentration at all temperatures[43]. At 400 K, a remarkable power factor of 554 μW/cm·K$^2$ is achieved in the p-type GAL with a square pattern, which is far beyond the best power factors of bulk thermoelectric materials[36].

In summary, representative GAL samples fabricated from CVD graphene were used to demonstrate a recently developed method[13] to analyze the scattering mechanisms of charge carriers, by detecting the effective CEStS. It is found that the pore-edge-trapped charges can be the major scattering mechanism in GALs. The ultra-high power factor obtained for these GALs, combined with their high thermal conductivities, can largely benefit their applications in device cooling. The power factor can be largely enhanced with a gate voltage, allowing active control of the thermoelectric properties during the device operation. Similar idea can also be found for PbSe nanowires[51]. The technique used in this work can be extended to general 2D materials for thermoelectric and electronic studies.

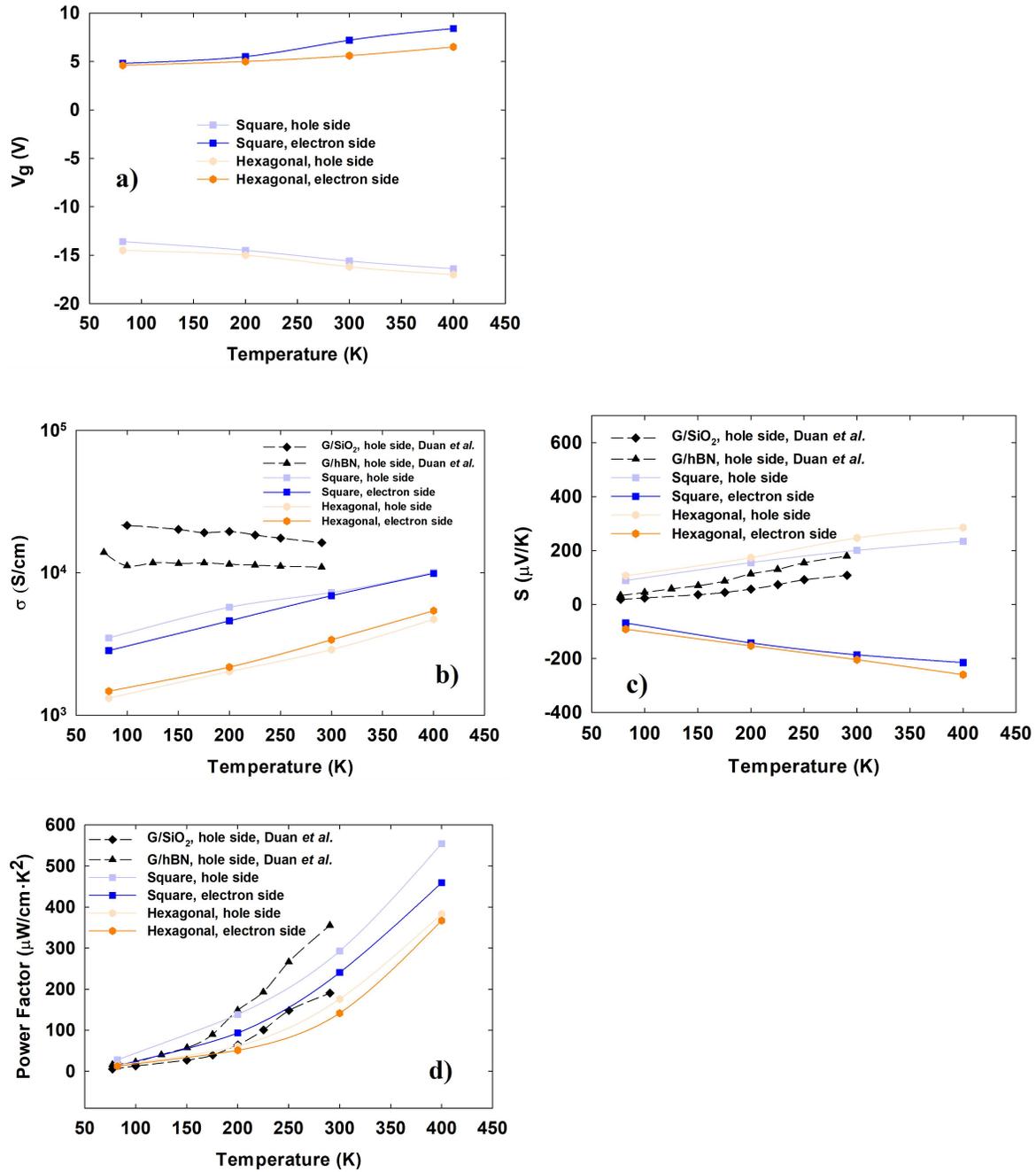

**Figure 7**. Temperature-dependent (a) optimized gate voltage for maximum $|S|$, (b) the corresponding electrical conductivity, (c) peak Seebeck coefficient, and (d) power factor of measured GALs. All results are compared to those measured for pristine graphene with a fixed gate voltage and thus carrier concentration[43].

ACKNOWLEDGE:

The portion of this work finished at the University of Arizona was supported by National Science Foundation CAREER Award (grant number CBET-1651840) for modeling of transport properties and AFOSR YIP Award (award number FA9550-16-1-0025) for studies on GALs.

COMPETING FINANCIAL INTERESTS:

The author(s) declare no competing interests.

AUTHOR CONTRIBUTIONS:

QH and ST designed the experiments, analyzed the data and wrote the paper. DX carried out all experimental measurements, and XD helped improve the experiments.